\documentclass[english,11pt]{article}

\usepackage{graphicx}
\usepackage{algorithm}
\usepackage{algorithmic}
\usepackage{amssymb}
\usepackage{amsmath}
\usepackage{amsxtra}
\usepackage{mathrsfs}
\usepackage{colortbl}
\usepackage{multirow}
\usepackage{color}

\makeatother

\begin{document}
\title{The Heisenberg spin glass model on GPU: myths and actual facts}

\author{M. Bernaschi$^\dagger$, G. Parisi$^{\diamond}$, L.
  Parisi$^\dagger$\\
$^\dagger$Istituto Applicazioni Calcolo, CNR, \\
Viale Manzoni, 30 - 00185 Rome, Italy \\
$^\diamond$Physics Department, University of Rome ``La Sapienza'', \\
P.le A. Moro, 2  - 00185 Rome, Italy
}

\maketitle

\begin{abstract}
We describe different implementations of the 3D Heisenberg spin glass
model for Graphics Processing Units (GPU). The results show that
the {\em fast} shared memory gives better performance with respect to the
{\em slow} global memory only if a multi-hit technique is used.

\end{abstract}


%

\section{Introduction\label{sec:intro}}
In spite of the availability of high performance multi-core systems
based on traditional architectures, there is recently a renewed
interest in floating point {\em accelerators} and {\em co-processors}
that can be defined as devices that carry out arithmetic operations
concurrently with or in place of the CPU. Among the solutions that
have received more attention from the high performance computing
community there are the NVIDIA Graphics Processing Units (GPU),
originally developed for video cards and graphics, since they are able
to support very demanding computational tasks. As a matter of fact,
astonishing results have been reported by using them for a number of
applications covering, among others, atomistic simulations,
fluid-dynamic solvers and option pricing. Simulations of statistical
mechanics systems based on Montecarlo techniques are another example
of applications that may benefit of the GPU computing capabilities.
In the present work we report the results obtained by following
different approaches for the implementation of a typical statistical mechanics
system: the classic Heisenberg spin glass model.

The paper is organized as follows: Section \ref{sec:spin} contains a
short introduction to the features of spin systems that are of
interest from the computational viewpoint; Section \ref{sec:gpu}
summarizes the main features of the GPUs used for the experiments;
Section \ref{sec:results} presents the performances obtained by using
a number of possible approaches for the implementation of the 3D
Heisenberg spin glass model; Section \ref{sec:conc} concludes with a
summary of the main results and a perspective about possible future
activities in this field.

\section{Spin systems\label{sec:spin}}
In statistical mechanics ``spin system'' indicates a broad class of models
used for the description of a number of physical phenomena. Although
apparently quite simple, spin systems are far from being trivial
to be studied and most of the times numerical simulations (very often based on
Montecarlo methods) are the only way to understand their behaviour. A spin system is
usually described by its Hamiltonian which has the following general form
\begin{equation}
H = -\sum_{i \ne j}J_{ij}\sigma_i\sigma_j
\label{equ:spinequ}
\end{equation}
The spins are defined on {\em lattice} which may have one, two, three
or even a higher number of dimensions. The sum in equation
\ref{equ:spinequ} runs usually on the first neighbors of each spin (2
in 1D, 4 in 2D and 6 in 3D). The spin $\sigma$ and the coupling $J$
may be either discrete or continuous and their values determine the
specific model. In the present work we focus on the Heisenberg spin glass
model where $\sigma_i$ is a 3-component vector such that
$\overrightarrow{\sigma_i} \in \mathbb{R}, |\sigma_i|=1$ and $J_{ij}$
is gaussian distributed with average value equal to $0$ and variance
equal to $1$. \\In a
3-dimensional system of size $L^3$, the contribution to the total
energy of the spin $\overrightarrow{\sigma_i}$ with coordinates
$x, y, z$ such that $i=x+y\times L+z\times L^2$ is
\begin{eqnarray}
J_{x+1,y,z}\overrightarrow{\sigma}_{x,y,z} \cdot
\overrightarrow{\sigma}_{x+1,y,z} & + & J_{x-1,y,z}\overrightarrow{\sigma}_{x,y,z} \cdot
\overrightarrow{\sigma}_{x-1,y,z} + \nonumber\\
J_{x,y+1,z}\overrightarrow{\sigma}_{x,y,z} \cdot
\overrightarrow{\sigma}_{x,y+1,z} & + & J_{x,y-1,z}\overrightarrow{\sigma}_{x,y,z} \cdot
\overrightarrow{\sigma}_{x,y-1,z}+ \nonumber\\
J_{x,y,z+1}\overrightarrow{\sigma}_{x,y,z} \cdot
\overrightarrow{\sigma}_{x,y,z+1} & + & J_{x,y,z-1}\overrightarrow{\sigma}_{x,y,z} \cdot
\overrightarrow{\sigma}_{x,y,z-1}\label{equ:staple}\end{eqnarray}
where $\cdot$ indicates the scalar product of two $\sigma$ vectors.
In most Montecarlo techniques used for the simulation of the Heisenberg
spin glass model (Metropolis, {\em Heat Bath}, {\em etc.}) it is
necessary to evaluate the expression in equation \ref{equ:staple} for
each spin. The main goal of the present work is to present several
approaches and to assess what is the
most effective scheme to compute this expression on a GPU. As a
consequence we are not going to address other issues, like the
generation of random numbers, even if we understand their importance
for an efficient GPU based simulation of spin systems, because they are already
faced in other studies \cite{randomXGPU}. Actually, other authors already
described efficient techniques for the simulation, on GPU, of spin systems ({\em
  e.g.}, \cite{IsingGPU} for the Ising model in 2D and 3D  and
\cite{Jose} for the three-dimensional
Heisenberg anisotropic model). However their
analysis appears somehow limited since they present results basically
for a single implementation whereas a GPU offers several alternatives
for an effective implementation that deserve to be considered and analyzed.

\section{Graphic Processing Unit  and CUDA\label{sec:gpu}}

In table \ref{table:hw} we report the key aspects of the three GPUs we used for our
numerical experiments:  a Tesla C1060, a Tesla C2050 and a GTX 480.
The C2050 and the GTX 480 are based on the latest architecture
(``Fermi'') recently introduced by NVIDIA.

\begin{table}[h]
\label{table:hw}
\begin{center}
\begin{tabular}{|l|r|r|r|}
\hline
GPU model                          & Tesla C1060 & Tesla C2050 & GTX 480 \\ \hline
Number of Multiprocessors          & 30    &   14   &   15     \\
Number of cores                    &   240 &   448  &   480     \\
Shared memory per block (in bytes) & 16384 & 49152  &  49152      \\
L1 Cache (in bytes)                &  N/A  & 16384  &  16384      \\
L2 Cache (in Kbytes)               &  N/A  & 768  &  768      \\
Number of registers per block      & 16384 & 32768  &  32768      \\
Max Number of thread per block     &  512  &  1024  &  1024      \\
Clock rate                         &  1.3 Ghz  &  1.15 Ghz   & 1.4
Ghz\\
Memory bandwidth                   &  102 GB/sec.  & 144 GB/sec. & 177
GB/sec. \\
Error Checking and Correction (ECC) & No   & Yes & No \\ \hline

\end{tabular}
\end{center}
\caption{Main features of the NVIDIA GPUs used for the experiments}
\end{table}

\begin{figure}
\begin{center}
\includegraphics[width=0.4\textwidth]{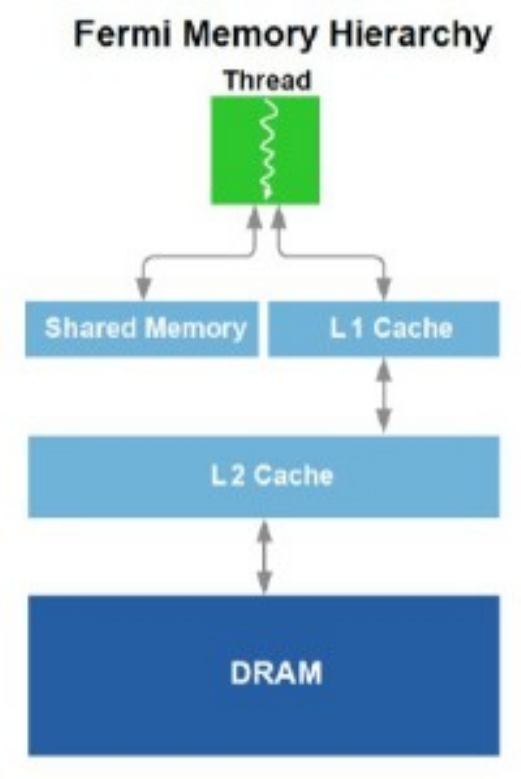}
\caption{Memory hierarchy in the Fermi architecture}
\label{fig:fermi}
\end{center}
\end{figure}

The memory hierarchy is one of the most distinguish features of the
NVIDIA GPUs and it includes:
\begin{itemize}
\item {\em global} memory (DRAM): this is the main memory of the GPU and any
  location of it is visible by any thread. The bandwidth between the
  {\em global} memory and the multiprocessors is more than 100 GB/sec
  but the latency for the access is also large (approximately 200
  cycles);
\item {\em shared} memory: access to data stored in the shared memory
  has a latency of only 2 clock cycles. However, shared memory variables
  are {\em local} to the threads running within a single
  multiprocessor and the size of the shared memory is tiny compared to
  the global memory that is, usually, in the range of GBytes;
\item {\em registers}: on a GPU there are thousands of 32 bits registers. It is worth noting that,
for each multiprocessor, there is more space for data in the
registers than in the shared memory.
\item {\em cache}: L1 and L2 caches have been included in the Fermi
  architecture. Actually, on each multiprocessor there are 64Kbytes of
  private L1 cache that can be split, at run time, in a 48 Kbytes shared memory and
  a 16KB L1 cache or in a 16Kbytes shared memory and a 48 L1 cache.
\item {\em constant} and {\em texture}: these are special memories
  used respectively to store constant values and to cache global
  memory (separate from register and shared memory) offering
  dedicated interpolation hardware separate from the thread
  processors.
\end{itemize}
Figure \ref{fig:fermi} summarizes the memory hierarchy of a GPU that
implements the Fermi architecture.

Data placement in the global or shared memory can be controlled
explicitly and this, as shown in Section \ref{sec:results}, makes a
significant difference from the performance viewpoint.\\
For the GPU programming, we employed the version 3.0 of the CUDA Software
Development Toolkit that offers an {\em extended} C compiler and is
available for all major platforms (Windows, Linux, Mac OS).
The extensions to the C language supported by the
compiler allow starting computational kernels on the GPU, copying data
back and forth from the CPU memory to the GPU
memory and explicitly managing the different types of
memory available on the GPU (with the notable exception of the caches)
The programming model is a Single Instruction Multiple Data (SIMD) type.
Each multiprocessor is able to perform the same operation on different
data 32 times so the basic computing unit (called {\em warp})
consists of 32 threads.
To ease the mapping of data to threads, the threads identifiers may be
multidimensional and, since a very high number of threads run in parallel, CUDA
groups threads in {\em blocks} and {\em grids}.\\
One of the crucial requirements to achieve a good performance on the
NVIDIA GPU is to hide the high latency of the {\em global} memory
accesses (both read and write) by following a set of rules that
depend on the specific level of the architecture (to achieve what is
called in CUDA jargon ``coalesced'' accesses). Also important is to
avoid running out of registers since registers-spilling, although
supported, has a very high cost.\\
Functions running on a GPU with CUDA have some limitations: they can
not be recursive; they do not support {\em static} variables; they do
not support variable number of arguments; function pointers are
meaningless.\\
Further information about the features of the NVIDIA GPU and the CUDA
programming technology can be found in \cite{CUDAProg}.

\section{Results\label{sec:results}}
\begin{figure}
\begin{center}
\includegraphics[width=0.8\textwidth]{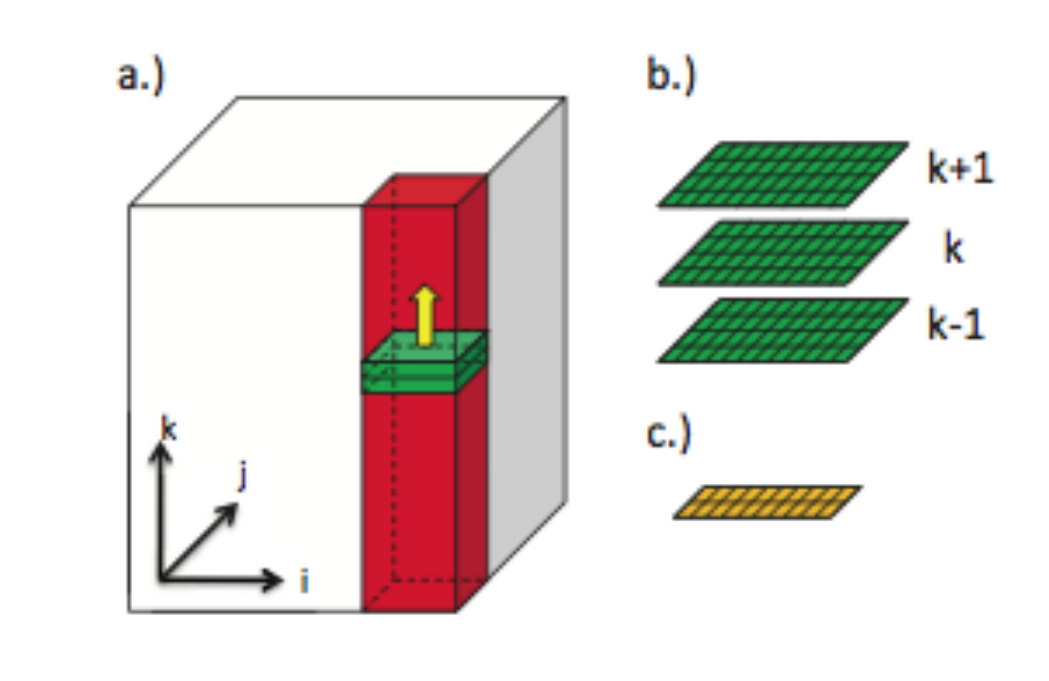}
\caption{The ``three planes'' mechanism}
\label{fig:3planes}
\end{center}
\end{figure}

For the tests we implemented, with different techniques, the so-called
Over-relaxation update \cite{Adler} in which for each spin
$\overrightarrow{\sigma}_{old} \rightarrow
\overrightarrow{\sigma}_{new}$ is the maximal move that leaves the
energy invariant, so that the change is always accepted. This dynamics
is micro-canonical however, it is very effective when used in
combination with an irreducible dynamics such as the {\em Heat Bath}.
The main reason for making this choice for our tests is that the
Over-relaxation update does not require random numbers. Moreover it requires
only simple floating point arithmetic operations (20 sums, 3 differences and
28 products) plus a single division.
Since there are no random numbers involved, the Over-relaxation update
can be hardly considered a Montecarlo method but nevertheless it is a
good benchmark since it requires the evaluation of the expression in
formula \ref{equ:staple} like real Montecarlo methods.
Likewise Montecarlo methods, the Over-relaxation update can be carried
out in parallel only on non-interacting spins. To this purpose a
check-board decomposition can be applied similar to that used for
vector processors \cite{DLandau}. This technique has been already implemented
for the GPUs in \cite{IsingGPU}. As already mentioned in Section
\ref{sec:gpu} an optimal data placement in the memory hierarchy of the
GPU is fundamental to achieve good performances. The shared memory of
the GPUs is very fast and it looks reasonable to use it for
storing spins and the coupling among them. Unfortunately, the total size of
the shared memory is limited (well below 1MB even on the latest
generation GPUs) and only very small systems can fit completely in it.
For the three dimensional Heisenberg model six memory locations per lattice
point are required (three for the components of the spin and three for
the couplings). In single precision, six memory locations occupy 24
bytes.
With a total size of the shared memory in the range of 0.5 Mbytes, only $\sim 22000$
lattice points could be stored corresponding to a linear size $L < 30$.
As a consequence to simulate systems with $L \ge 32$ a ``swapping''
mechanism is required. A similar problem arises trying to use a GPU to
solve a Laplace equation (the Laplace equation may be solved by using
the Jacobi scheme that requires the evaluation of an expression very similar to
that shown in formula \ref{equ:staple}). In this context a quite
elegant solution has been proposed in \cite{Giles} and \cite{Fatica}
where only three planes are stored in shared memory. The three planes
serve as a cyclic buffer. At each step along the $Z$-direction  a new $XY$
plane enters in the buffer replacing that plane that does not serve
to compute the output for the current plane. Such scheme is shown in
figure \ref{fig:3planes}. However, this scheme when applied in such a simple
way to a spin system has a major drawback since on each plane only
half of the spins can be updated concurrently (due to the constraint
of updating only non-interacting spins). As a consequence, half of
the threads are idle waiting their turn. To overcome this limitation,
we developed a new scheme which makes use of four planes instead of
three. In this scheme two consecutive planes (say the planes $k$ and
$k+1$) are updated concurrently in two sub-steps:
\begin{itemize}
\item {\em sub-step 1}: the white spins of plane $k$ and the black spins of
  plane $k+1$;
\item {\em sub-step 2}: the black spins of plane $k$ and the white spins of
  plane $k+1$.
\end{itemize}
In this scheme two planes are replaced at the end of each step along
the $Z$ direction and each step increases $Z$ of 2 units. A multi-hit
variant of this four-planes scheme has been developed as well. The
multi-hit version allows to measure the effect of the initial loading.\\
For the Fermi architecture a further scheme has been implemented to
measure the advantage provided by the cache. This scheme is quite
similar to the three-plane shared memory scheme, meaning that a single
plane is updated by changing concurrently all white spins and then all
black spins. Finally, a version where the loading of data from the {\em
  global} memory is replaced with texture fetches has been also
developed. Texture memory provides cached read-only access that is
optimized for spatial locality and it should prevent redundant loads
of global memory. When several blocks request the same region, the
data are loaded from the cache. We wanted to test whether the texture
helps with a memory access pattern like that required for the evaluation of
the expression in formula \ref{equ:staple}.\\
All the tests have been carried out for a cubic lattice with periodic
boundary conditions along the $X, Y$ and $Z$ directions. The indexes
for the access to the data required for the evaluation of the
expression \ref{equ:staple} are computed in accordance with the
assumption that the linear size $L$ of the lattice is a power of 2. In
this way bitwise operations and additions suffice to compute
the required indexes with no multiplications, modules or other costly
operations. Other details about the implementation of the different
approaches can be found looking directly at the source code available from\\
\verb|http://www.iac.rm.cnr.it/~massimo/hsgfiles.tgz|. Most of the
tests have been carried out on a lattice with linear size $L=128$.
The time we report is in nanoseconds and corresponds to the time
required to update a single spin. All the calculations are done in
single precision. The correctness of the algorithms is confirmed by
the fact that the energy remains the same (as expected since the dynamics of
the over-relaxation process is micro-canonical) even if the spin
configuration changes step-by-step. To have a reference point on a
standard architecture we implemented also a highly tuned CPU version for
an Intel {\em i7} multicore with a cache of
8MBytes running at 2.93 Ghz. The CPU version makes use of the
vector instructions (SSE) of the Intel architecture and is parallelized by
using the OpenMP directives.

\begin{table}[h]
\begin{center}
\begin{tabular}{|l|r|r|}
\hline
{\em Platform}  & {\em number of threads} &  $T_{upd}$ \\ \hline
Tesla C1060 GM & 320 & 1.9 ns \\
Tesla C1060 CA & 320 & 2.0 ns \\
Tesla C1060 SM & self determined & 2.5 ns\\
Tesla C1060 SM4P & self determined & 2.2 ns \\
Tesla C1060 TEXT & 320  & 1.8 ns \\ \hline
GTX 480 (Fermi) GM & 320 & 0.66 ns\\
GTX 480 (Fermi) CA & 320 & 0.70 ns\\
GTX 480 (Fermi) SM & self determined & 1.3 ns  \\
GTX 480 (Fermi) SM4P & self determined & 0.86 ns \\
GTX 480 (Fermi) TEXT & 480 & 0.63 ns \\ \hline
Intel {\em i7} (SSE instr.) & 1 & $\sim 13.5$ ns  \\
Intel {\em i7} (SSE instr. + OpenMP) & 8 & $\sim 5$ ns \\ \hline
\end{tabular}
\end{center}
\caption{Timings for simulating a single Heisenberg spin glass system with
continuous (gaussian distributed) isotropic couplings.
The lattice size is set equal to $L=128$. $T_{upd}$ is the time in
nanoseconds to process 1 spin. We ran 100 iterations and report the
total time divided by the number of iterations and then divided by the
number of spins in the system}\label{tab:results}
\end{table}

The main results are reported in table \ref{tab:results}. GM stands
for Global GPU Memory. This is the most simple version which starts a
very large number of threads and blocks by exploiting the fact that
the Overrelaxation GPU kernel we wrote needs very few registers.
The GPU kernel updates all white spins and then all black spins.
The synchronization required between the updating of white and black
spins is guaranteed by two distinct kernel invocations.
To reduce the corresponding overhead, we could resort to a different
synchronization mechanism although the lack of a native mechanism to
synchronize, within a single kernel, the thread-blocks each other
makes a bit tricky any alternative approach.

SM stands for Shared GPU Memory. This is the version which keeps three
planes in shared memory and updates before white and then black spins of a
plane before proceeding to the next plan. SM4P stands for Shared
Memory version with 4 planes in memory. The difference with the SM
version is that we update two
plane concurrently (the white spins on the first plane and the black
ones on the second plane and then the other way around: the black
spins on the first plane and the white ones on the second plane).
For the two shared memory versions (SM and SM4P) the number of threads
is determined implicitly by the size of the available shared memory
(so it varies between the Tesla C1060 where it is equal to 144 and the
GTX 480 or the C2050 where it is equal to 384).

CA stands for ``Cache''. This is the version developed for the Fermi
architecture where a 48Kbyte cache is available on each
MultiProcessor. The results in the table have been obtained by using
the {\em hint} for L1 cache that sets it equal to 48K per MP. The idea
is to update all the white spins of a plane and then all the black
spins of the same plane. If the cache works as expected, three planes
should be loaded when the white spins are updated. Then for the update
of the black spins no new data should be loaded. We tested this
version also on the Tesla architecture, although the Tesla does not
have a cache, just to measure the overhead introduced by processing a
single plane in each kernel.\\
Finally, we measured the {\em overhead} of invoking a CUDA kernel. This
is basically independent on the method and represents a lower
bound of the time that the update of a spin requires. The value we
found is $0.1$ ns per spin.

\begin{table}[h]
\begin{center}
\begin{tabular}{|r|r|r|}
\hline
{\em Number of hits}  & $T_{upd}$ {\em sm\_arch=20} & $T_{upd}$ {\em sm\_arch=13}\\ \hline
 1 & 0.911 ns &  0.856 ns\\
 2 & 0.606 ns & 0.546 ns\\
 5 & 0.429 ns & 0.366 ns  \\
10 & 0.373 ns &  0.311 ns \\
$\infty$ (extrapolated) & 0.313 ns & 0.246 ns\\ \hline
\end{tabular}
\end{center}
\caption{Timings with respect to the number of hits for the SM4P
  version on a GTX 480. Timings are reported for the compilation with
  two different setups (arch={\em sm\_20}, corresponding to the latest
  generation software that works only on the Fermi architecture and
  arch={\em sm\_13} that works for both the previous and the latest architecture). The
test case is the same as in table \ref{tab:results}. \label{tab:hits}}
\end{table}

\begin{table}[h]
\label{table:hsg-timings}
\begin{center}
\begin{tabular}{|l|r|r|}
\hline
{\em Platform}  & $T_{upd}$ {\em ECC on} & $T_{upd}$ {\em ECC off}\\ \hline
C2050 (Fermi) GM & 1.0 ns &  0.84 ns\\
C2050 (Fermi) CA & 1.1 ns & 0.87 ns\\
C2050 (Fermi) SM & 1.63 ns & 1.66 ns  \\
C2050 (Fermi) SM4P & 1.4 ns & 1.3 ns \\
C2050 (Fermi) TEXT & 1.0 ns &  0.78 ns \\ \hline
\end{tabular}
\end{center}
\caption{Timings with and without Error Checking and Correction on a
  C2050 (the GTX 480 does not have ECC). The
test case is the same as in table \ref{tab:results}. The compilation
setup is arch={\em sm\_13}. \label{tab:ecc}}
\end{table}

\section{Discussion and Conclusions\label{sec:conc}}
In table \ref{tab:hits} we report the results of a test in which the
spins of the version SM4P are updated multiple times (``multi hit'')
before moving  to the next two planes. Since the updates after the
first one do not require new accesses to the global memory (all data
are already in the shared memory) this technique allows to measure the
overhead introduced by the initial loading of the data as the
difference between the total time required by the single hit and the
time required by an infinite number of hits. Obviously it is not
possible to carry out an infinite number of hits but it is easy to
extrapolate the corresponding value by using a simple fit. The time we
found is $\sim 0.6$ ns for both the compilation setups used for the
tests. Interestingly, at this time, it is more efficient to compile
on the Fermi architecture with a setup ({\em arch=sm\_13}) introduced
for the previous generation GPUs. On the other hand it is absolutely
necessary to take into account the specific features of the new
architecture like the different size of the shared memory or the new
integer multiplication capability. For instance, if on the Fermi
architecture the multiplication between two integers is carried out by
using the \verb|__mul24| (as it was suggested on the previous
generation GPUs when the result of the multiplication fitted in 24
bits) there is a penalty of about $5\%$ for our code.
If we consider the time ($0.63$ ns per spin update) required by the
best ``single-hit'' technique, that is the texture based one, and use
the estimate of 60 floating point operations per spin update (that
does not take into account the index arithmetics), we obtain a
sustained performance very close to $100$ GFlops. Besides that,
the most interesting observation is that, despite the much higher
latency of the global memory, the simple versions (the GM one and the
TEXT one) that use neither the shared memory nor the cache (where
available) are significantly faster unless a ``multi-hit'' variant of
the shared memory version is used. A possible explanation of this
behaviour is that the limited size of the shared memory (and of the
cache) allows to start fewer threads with respect to the case of a
global memory based version where the only limitation is the number of
registers. Moreover, although the shared memory version allows to reduce
the number of global memory transactions when loading data (spins and
couplings), it offers no advantage when the updated spins are stored
back in global memory. The advantage of using texture fetches instead
of simple load operations from the global memory is quite limited (about $5\%$).
However, only very minor changes are required to the code to take this
advantage so it is worth it. It is more puzzling that, apparently the
cache does not offer advantages at all. A possible explanation is that
the overhead introduced by calling the computational kernel a number
of times equal to twice the linear dimension of the lattice (instead of
calling it just once as we do in the GM or TEXT version) is as large
and possibly larger than the saving in time that the cache may offer.
The requirement of calling the kernel multiple times arises because on
each plane it is necessary to update all white spins and then all
black spins. So, for each step in the $Z$ direction, two invocations
are required. \\
As shown in table \ref{tab:ecc}, the Error Checking and Correction
introduces a sizeable overhead for the methods working in global
memory (GM, TEXT, CA). For reasons that we could not determine, the
methods working in shared memory undergo a much more limited penalty
when the ECC is on.\\
Finally, it is worth to highlight two facts: {\em i)} a well tuned GPU implementation can achieve
almost an order of magnitude of speedup with respect to a well tuned
vector-multicore  implementation ($0.63$ ns per spin update for the TEXT version on the
GTX 480 {\em vs.} $5$ ns per spin update on a multicore Intel {\em
  i7}; {\em ii)} the performances of the GPUs keep on to increase significantly
in a pretty short time (the performances of the new architecture are
double and in some cases even more than double with respect to the
previous generation cards that are only two years old)
making it a very interesting platform for the numerical simulation of
spin systems.

\section*{Acknowledgments}
We thank Luis Antonio Fernandez, Filippo Mantovani, Victor
Martin-Mayor, Sergio Perez, Fabio Schifano and Raffaele Tripiccione
for useful discussions.\\
We thank Massimiliano Fatica for the {\em hint} about the
three planes technique and for running preliminary tests on Fermi
architecture based GPUs.\\
We thank Luca Leuzzi and Edmondo Silvestri for the access to their GTX
480 GPU.\\
Special thanks to Jos\'e Manuel Sanz Gonz\'alez for providing us with
his program for simulating the Heisenberg spin glass (including
anisotropic couplings) and sharing his preliminary performance results.

\end{document}